\documentclass[twocolumn,prb,showpacs]{revtex4-1}
\usepackage{times,xspace}
\usepackage{graphicx,graphics,color,epsfig}
\usepackage{bm}
\usepackage{amsmath}
\usepackage{amssymb}
\usepackage{epstopdf}
\usepackage{dcolumn}

\begin{document}

\title{Visualizing electron pockets in cuprate superconductors}

\author{Tanmoy Das$^1$, R. S. Markiewicz$^2$, A. Bansil$^2$, and A. V. Balatsky$^{1,3}$}
\affiliation{$^1$Theoretical Division, Los Alamos National Laboratory, Los Alamos, NM, 87545, USA.\\
$^2$Physics Department, Northeastern University, Boston MA 02115, USA\\
$^3$Center for Integrated Nanotechnologies, Los Alamos National Laboratory, Los Alamos, NM, 87545, USA.}
\date{\today}

\begin{abstract}
Fingerprint of the electron-pocket in cuprates has been obtained only in numerous magneto-transport measurements, but its absence in spectroscopic observations pose a long-standing mystery. We develop a theoretical tool to provide ways to detect electron-pockets via numerous spectroscopies including  scanning tunneling microscopy (STM) spectra, inelastic neutron scattering (INS), and angle-resolved photoemission spectroscopy (ARPES). We show that the quasiparticle-interference (QPI) pattern, measured by STM,   shows additional 7 ${\bm q}$ vectors associated with the scattering on the electron-pocket, than that on the hole-pocket. Furthermore, the Bogolyubov quasiparticle scatterings of the electron pocket may lead to a second magnetic resonance mode in the INS spectra at a higher resonance energy. Finally, we reanalyze some STM, INS, and ARPES experimental data of several cuprate compounds which dictates the direct fingerprints of electron pockets in these systems.
\end{abstract}

\pacs{74.25.Jb,74.72.Gh,74.72.Kf} \maketitle\narrowtext
Copper-oxide high-temperature superconductors evolve from a Mott insulator to the superconducting state  through an unknown `pseudogap' phase. Many competing order origins of the `pseudogap' have been proposed, some of which lead to a Fermi surface (FS) reconstruction into hole-pocket and electron-pockets.\cite{Chakravarty,Norman_stripe,Sachdev,LaliberteNernst,Sachdev_Nernst} Hole-pockets are detected in many experiments.  On the other hand the existence of electron pockets has been overlooked for the past twenty years and only recently has been proposed by Hall-effect, quantum oscillation at high magnetic field, Nernst and Seebeck measurements.\cite{Taillefer,TailleferQO,Sebastian,LaliberteNernst}
In particular, Hall-effect measurements have revealed a negative sign in the low-temperature Hall coefficients which is taken as a signature of electron-like quasiparticles on the FS.\cite{Taillefer} The Hall-coefficient in fact changes sign from negative to positive with increasing temperature but below $T^*$, suggesting the coexistence of both electron and hole-pockets on the FS. Shubnikov-de-Haas (SdH) experiments in YBa$_2$Cu$_3$O$_{6.5}$ and YBa$_2$Cu$_4$O$_8$ (YBCO) also argue for the presence of closed FS pockets,\cite{footEP} with slope suggestive of electron-pockets.\cite{TailleferQO,Sebastian} This observation received further supports from the Nernst and Seebeck measurements which have been shown theoretically to be consistent with the coexistence of electron and hole-pockets.\cite{LaliberteNernst,Sachdev_Nernst}
The question arises, {\it if an electron pocket is present on the FS, are there spectroscopic fingerprints that can detect it directly?} For example, ARPES which directly measures the single-particle spectral weight, has so far been unable to convincingly separate out the presence of an electron-pocket from a full paramagnetic FS.

Many theoretical proposals have been put forward to explain the FS topology in cuprates,\cite{Chakravarty,Norman_stripe,Sachdev} however, a consistent picture to describe both the bulk measurements and the spectroscopies has yet not been achieved.  Within a strong coupling scenario, the holes, doped into the parent Mott insulator, create in-gap states at the Fermi level without a well defined quasiparticle dispersion.\cite{PLee} Again in the pre-formed SC pairing theory of the `pseudogap', one would predict that a single large hole-like FS persists at all dopings, with SC fluctuations suppressing spectral weight in the antinodal regions, leaving a Fermi arc.\cite{Norman_preformed} Such a model would predict a hole-like sign of the Hall coefficient at all temperatures,\cite{QOFA} incompatible with the observed sign changing Hall-effect\cite{TailleferQO,Sebastian} and Nernst and Seebeck measurements.\cite{LaliberteNernst} An alternative approach using a density wave picture of the pseudogap has been successful in explaining many aspects like the behavior of quantum oscillations, Hall, Nernst and Seebeck effects,\cite{Norman_stripe,Sachdev_Nernst,Chakravarty,Harrison} and ARPES, STM and neutron scattering.\cite{Dastwogapcuprate,Dasop,Dasresonance} Of course, in YBCO there are other band-structure properties that can serve as electron-like FS.\cite{DasEP}

To find signatures of electron pockets, we model the pseudogap as a spin-density wave (SDW) state which leads to the FS reconstruction into hole and electron pockets.\cite{Dastwogapcuprate}. Using this model, we find that (i) the QPI pattern seen in STM exhibits 7 new ${\bm q}-$vectors which evolve in a qualitatively different way than the ones expected for a hole-pocket; (ii) similarly, the INS measurements also display an additional resonance peak in the spin-excitation spectrum in the SC state coming from the electron-pocket; (iii) furthermore, in some doping regions ARPES FS spectral weight data reveal two peaks at the nodal and antinodal points with a dip between them which suggests reconstruction of the FS into hole and electron pockets, respectively; (iv) we also  demonstrate several key properties of these three spectroscopies which quantitatively and unambiguously can establish the presence of an electron-pocket on the FS.

The development of electron pocket in hole doped cuprates is doping (and material) dependent. In the overdoped region, the FS consists of a large hole-like FS centered at $M=(\pi,\pi)$. At strong underdoping, a pseudogap opens in the region of momentum space near ${\bm k}=(\pi,0)$ and $(0,\pi)$, leaving a hole-pocket or `Fermi arc' at the nodal point ${\bm k}=(\pi,\pi)$. These hole pockets are observed directly by ARPES and are consistent with STM and many other experiments. With increasing doping, as the pseudogap correlation weakens but remains finite, the bottom of the conduction band at ${\bm k}=(\pi,0)$ and $(0,\pi)$ drops down below the Fermi level producing an electron pocket at some critical doping, even without the application of any external magnetic field (see Appendix~\ref{App1} for the details of the evolution of the FS). A small gap persists in the regions where the bare FS crosses the magnetic Brillouin zone [marked by a dashed line in Fig.~1(a)]. As the electron-pockets are expected to form in the doping range where the FS crosses over from small pocket to large FS, spectroscopies need guidance to distinguish a pocket from a full  hole-like FS. Therefore,  we provide a careful analysis of the spectroscopic details to illustrate how to observe the electron-pocket.

\begin{figure}[top]
\hspace{-0cm}
\rotatebox{0}{\scalebox{.38}{\includegraphics{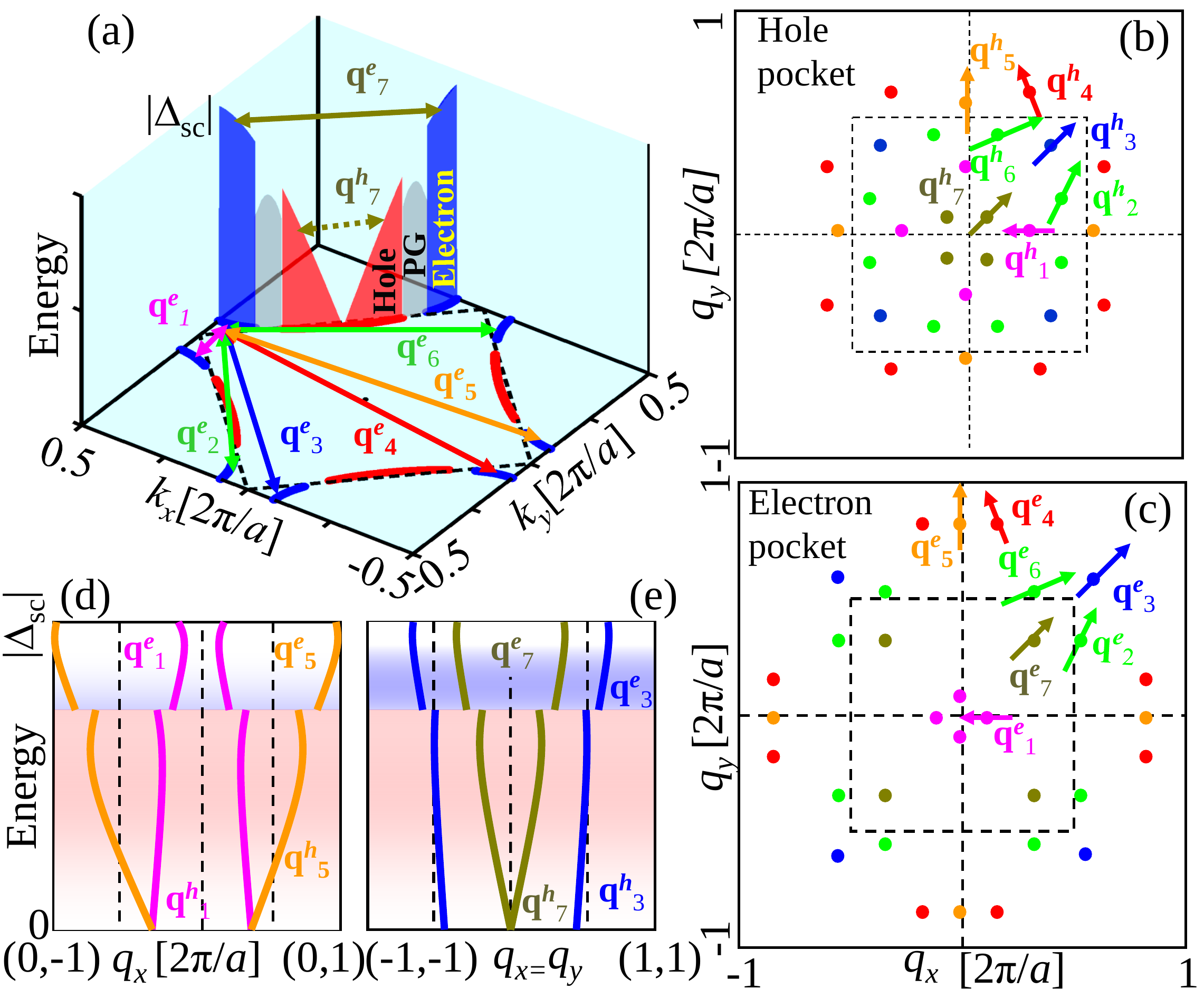}}}
\caption{{Schematic QPI pattern for electron pockets.} (a) Sketch of hole-pockets (red lines) and electron-pockets (blue lines). The front sides  of the two pockets (main bands) are drawn here, where the induced shadow bands are not shown. Opening of a $d-$wave SC gap on these pockets is shown in color shadings in one quadrant of the FS. The seven QPI vectors connecting eight elastic bright spots on a constant energy surface on the electron pocket are shown by arrows of various colors. The contrast between the QPI vectors associated with the hole-pocket and the electron-pocket is illustrated for one vector ${\bm q}_7^{h,e}$ only, while the same for other vectors follows similarly. (b) A view of a constant energy QPI map of hole pocket origin is contrasted with the same from an electron pocket origin in (c). Arrows of same color point to the direction of the motion of each ${\bm q}-$vector with increasing energy. (d)-(e) The dispersive behavior of the QPI vectors in the ${\bm q}-\omega$ phase space is schematically shown along the high-symmetry lines of (100)-direction in (d) and along the diagonal direction in (e). The red and blue background shadings differentiate the hole pocket and electron pocket regions. All the QPI vectors show kinks in going from the hole pocket to the electron pocket energy which is an indicator of the presence of the electron pocket on the FS.}
\label{fig1}
\end{figure}

In the superconducting state, the $d-$wave pairing restricts the coherent Bogoliubov quasiparticles to move on the ${\bm k}-$space of the electron and hole pockets, see Fig.~\ref{fig1}(a). The scattering process of these particle-hole excitations leads to many observable features, like the elastic scatterings of the Cooper pairs seen as a QPI pattern in STM.\cite{Hanaguri} Similarly, inelastic scattering between particle and hole Bogoliubov quasiparticles leads to a scattering profile as revealed by INS.\cite{chubukov} The QPI and INS patterns generated by the hole-pocket are well studied in cuprates.\cite{norman,eremin} Here we study how these patterns evolve naturally to include contributions of the electron pocket.

{\bf Scanning Tunneling Microscopy$-$} Figures.~\ref{fig1}(b) and \ref{fig1}(c) contrast the QPI patterns at two representative quasiparticle energies at which the Cooper pair resides on the hole-pocket (lower energy) and electron pocket (higher energy), respectively. There is a qualitative difference in the overall QPI pattern at these two energy scales. First, since scattering is purely elastic, appearance of an electron pocket leads to new features in QPI that correspond to 7 additional electron-electron scattering ${\bm q}^e$ vectors in addition to hole-hole scattering ${\bm q}^h$ vectors. No elastic scattering features connect electron and hole pockets as they have different quasiparticle energies. The definitive distinction between the two pockets can be marked by the values of two high symmetry vectors ${\bm q}^{h,e}_3$ and ${\bm q}^{h,e}_5$. ${\bm q}^{h}_3$ connects equivalent energy points on two hole-pockets along the diagonal direction. As the hole-pocket terminates at the magnetic zone boundary at which ${\bm q}^{h}_3=(\pi,\pi)$, therefore, if ${\bm q}^h_3$ continues to grow above $(\pi,\pi)$, it must come from the electron pocket. Similarly, ${\bm q}^{h}_5$ [along the (100)-direction] will attain its maximum value equal to the reciprocal lattice vector of $(2\pi,0)$ and $(0,2\pi)$ at the highest energy of the QPI pattern.

In addtion, one requires to pay attention to the energy dependence of the QPI vectors as well as their associated intensities. Due to the van-Hove singularity at the antinodal point as well as the discontinuous jump from the hole-pocket to the electron-pocket FS, one expects a `kink' in the energy dependence of each QPI vector, see Figs.~\ref{fig1}(d) and \ref{fig1}(e). As the ${\bm q}^h$ vectors reach the top of the hole-pocket [i.e. when ${\bm q}_3=(\pi,\pi)$], the Bogoluybov scattering of these vectors vanishes and they become merely FS nesting. Therefore, all ${\bm q}^h$ vectors shoot almost vertically upward but with diminishing intensity. Nearly at the same energy, the Bogolyubov scattering on the electron pocket turns on and ${\bm q}^e$ vectors appear on the QPI pattern. Unlike  ${\bm q}^h$s, ${\bm q}^e$s disperse slowly with energy but the associated intensity begins to rise again. Therefore, not only the magnitude of the ${\bm q}^e$ vectors as discussed above, but also the expected `kink' in their dispersion and their associated intensity will serve as quantitative and unambiguous marks for the presence of electron-pockets.

\begin{figure*}[top]
\rotatebox{0}{\scalebox{.7}{\includegraphics{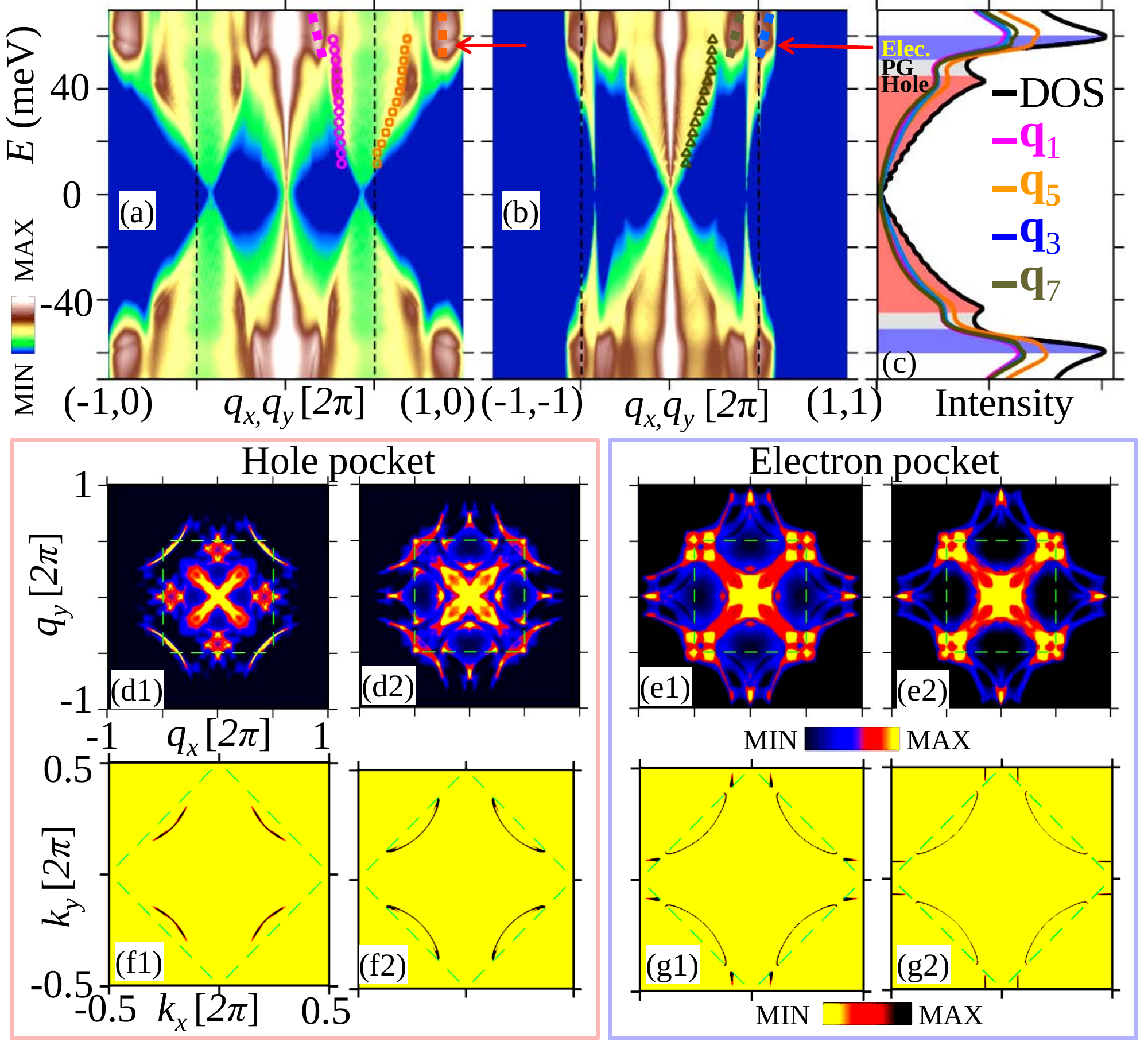}}}
\caption{{Computed QPI pattern due to the electron-pocket.} (a)-(b) The momentum-energy dispersion relation of the QPI pattern is drawn along (100)-direction and the diagonal one, respectively. In these two high-symmetry directions only four QPI ${\bm q}$ vectors appear as highlighted by dashed lines. The dots are the experimental data of Bi2212 in an overdoped sample $T_c=75$K for the same ${\bm q}$ vectors generated from the hole-pocket, plotted only in one direction for clarity.\cite{McElroy} These experimental data are shifted along the ${\bm q}$-directions by $\Delta q=0.08(2\pi)$ to reconcile the fact the FS areas for Bi2212 and YBCO (theory) are different and the energy axis is scaled by $\Delta_{YBCO}/\Delta_{Bi2212}=2.23$, where $\Delta$ is the SC gap. At the termination of the hole-pocket both the experiment and theory consistently reproduce the non-dispersive nature of the hole-pocket QPI vectors. The QPI vectors from the electron pocket appear in this energy region. (c) Theoretical DOS, black line, and the intensities of various QPI vectors (see legend) exhibit a one-to-one correspondence with each other. All the spectra exhibit linear-in-energy dependence coming from the $d-$wave nature of the SC gap and have two characteristic peaks at the tip of the hole-pocket (low-energy peak) and at the tip of the electron-pocket. Computed QPI patterns in the two-dimensional momentum space at four energy values; (d1) and (d2) correspond to the hole-pocket while (e1) and (e2) are obtained in the electron-pocket region. (f1)-(f2) and (g1)-(g2) The single-particle maps of `bright-spots' in the $k-$space of the Bogolyubov quasiparticle are plotted at the same energy values at which the QPI maps are calculated in the corresponding upper panel.}
 \label{fig2}
\end{figure*}

To demonstrate how the electron-pocket leads to a different set of QPI patterns, we calculate the QPI spectra in a coexisting uniform phase of SDW induced pseudogap and $d-$wave superconductivity.\cite{Dastwogapcuprate} We concentrate on YBCO$_{6.6}$ where the band dispersion is obtained by the tight-binding fitting to the first-principle calculations. Based on this ground state, the self-energy correction due to spin and charge fluctuations is computed within a self-consistent $GW$-model [see Appendix~\ref{App1}].\cite{Dasop} The lifetime broadening due to the imaginary part of the self-energy helps create `bright-spots' on the constant energy single-particle spectra. At any energy in the SC state, we have 8 `bright spots' due to $d-$wave symmetry as shown in Figs.~\ref{fig2}(f1),~\ref{fig2}(f2),~\ref{fig2}(g1),~\ref{fig2}(g2) at four representative energy cuts below the Fermi level. At $E=0$, the `bright-spots' are concentrated at the nodal points (not shown) and with increasing energy, they move towards the antinodal direction. The locus of the `bright-spot' is always restricted to move on the normal state FS and takes the form of well-known `banana-shape' in the low-energy region (on the `hole-pocket'); see Figs.~\ref{fig2}(f1) and \ref{fig2}(f2). As the `bright-spots' hit the magnetic zone boundary [green dashed line], they move to the electron-pocket region, see Figs.~\ref{fig2}(g1) and \ref{fig2}(g2).

We calculate the QPI pattern as a convolution of the constant energy single particle spectra $B({\bm q},\omega)\sim\sum_{\bm k}{\rm Im}\bigl[G({\bm k},\omega)G({\bm k}+{\bm q},\omega)\bigr]$, where $G$ is the $4\times 4$ single particle Green's function in the SDW-SC state, see Appendix~\ref{AppQPI}. At $E=0$, ${\bm q}^h_1$ and ${\bm q}^h_5$ are the same vector connecting the nodal points. With increasing $|E|$, ${\bm q}^h_1$ gradually shrinks whereas ${\bm q}^h_5$ grows$-$both very much linearly with energy, coming from the linear dispersion of the nodal Bogoliubov quasiparticles. A similar linear dispersion is evident in the behavior of ${\bm q}^h_3,{\bm q}^h_7$. ${\bm q}^h_7$ starts from ${\bm q}=0$ at $E=0$ and increases to a maximum value less than $(\pi,\pi)$ in all underdoped cuprates while ${\bm q}_3$ starts at a finite vector slightly below $(\pi,\pi)$ and reaches $(\pi,\pi)$ at the edge of the hole pocket. The resulting two dimensional QPI pattern in this energy scale is shown in Figs.~\ref{fig2}(d1) and \ref{fig2}(d2), and agrees qualitatively with the experimental results of Bi2212.\cite{McElroy} Above this energy, all ${\bm q}^h$ vectors become normal-state FS nesting,  which is not related to the Bogoliybov scattering, and bend backward with much less dispersion while the associated intensity gradually diminishes. Therefore, in the absence of an electron pocket, one can expect the QPI pattern to remain very much same as a function of energy but with much broadened peaks due to the lack of Bogolyubov coherence peaks. The other weak-intensities apart from the leading 7 ${\bm q}$-vectors are associated with the shadow bands, which are not relevant for the present study.

The most interesting feature of the QPI happens above the pseudogap energy scale which separates the electron pocket from the hole pocket. New ${\bm q}-$vectors develop due to the Bogoluibov scattering of the electron pocket. These ${\bm q}^e$ vectors are practically the continuation of the ${\bm q}^h$s above the magnetic zone boundary but with different slope and intensity which are related to the curvature of the electron pocket and the associated van-Hove singularity. The resulting constant energy QPI maps are shown in Figs.~\ref{fig2}(e1)-\ref{fig2}(e2) with very distinct interference patterns compared to the hole pocket [compare with Figs.~\ref{fig2}(d1) and \ref{fig2}(d2), respectively]. In the electron pocket regions, only ${\bm q}^e_1$ disperses toward ${\bm q}=0$, whereas the others disperse away from the magnetic zone boundary to the reciprocal unit cell boundary. ${\bm q}^e_3,q_7^e,q^e_2,q_6^e$ approach each other forming a squarish profile centered at ${\bm q}=(\pi,\pi)$ which is present at all energies. Also, ${\bm q}^e_2,q_5^e$ approach ${\bm q}=(2\pi,0)$ and its equivalent $k$-points. We emphasize that the most robust features signaling the presence of the electron pocket will be the values of ${\bm q}^e_3$, ${\bm q}^e_2,q^e_5$ in that ${\bm q}^e_3>(\pi,\pi)$ at the beginning of electron pocket whereas ${\bm q}^e_2,q^e_5$ reach the zone boundary ($2\pi,0)$ [and its equivalent points] at the top of the electron-pocket.

The intensity of each ${\bm q}^h$ and ${\bm q}^e$ vector follows closely to the density of states (DOSs) as shown in Fig.~\ref{fig2}(c). Both the DOS and the QPI intensity grow linearly with $|E|$, demonstrating $d$-wave pairing symmetry and the particle-hole symmetry in the Bogoliubov quasiparticles even in the pseudogap state and also under the influence of many-body effects. Above the tip of the hole-pocket, the intensity drops in the pseudogap energy region and then it rises again sharply up to the tip of the electron pocket. Experimentally the first peak in intensity is well documented for underdoped samples while some evidence of the second peak is seen in overdoped Bi2212 [see for example Ref.~\onlinecite{JLee}].

We summarize three robust signatures that help unambiguously differentiate the presence of the electron pocket from the hole pocket or paramagnetic full FS (see appendix~\ref{App3} for details). (1) In the dispersion relation of the QPI vectors as shown in Figs.~\ref{fig2}(a) and \ref{fig2}(b), all the ${\bm q}$ vectors stop dispersing at the tip of the hole-pocket. Only for the case of an electron pocket, new QPI vectors appear which extend to ${\bm q}=(2\pi,0)$ and its equivalent points along the (100)-direction or above ${\bm q}=(\pi,\pi)$ along the diagonal direction. Furthermore, to differentiate an electron pocket from a paramagnetic full FS, one needs to pay attention to the break in the slope of the QPI vectors going from the hole-pocket to the electron pocket. (2) For constant energy scans, the QPI profile becomes essentially energy independent above the tip of the hole-pocket; for example, when ${\bm q}^h_3=(\pi,\pi)$ stops dispersing. In contrast, in the present case of an electron-pocket, the new QPI pattern forms with two distinguishing marks that ${\bm q}^e_3>(\pi,\pi)$ and ${\bm q}^e_5=(2\pi,0)$ at the tip of the electron-pocket. (3) The intensity of the QPI vectors as a function of energy shows two distinct peaks in the case when both electron and hole pocket are present on the FS. Lower energy peak occurs at the tip of the hole pocket at an energy $|E|<|\Delta|$ while the second peak happens at the tip of the electron pocket exactly at $|E|=|\Delta|$. In the absence of an electron pocket, only the first peak will be present whereas in a paramagnetic ground state only the second peak will show up.

\begin{figure*}[top]
\hspace{-0cm}
\rotatebox{0}{\scalebox{.6}{\includegraphics{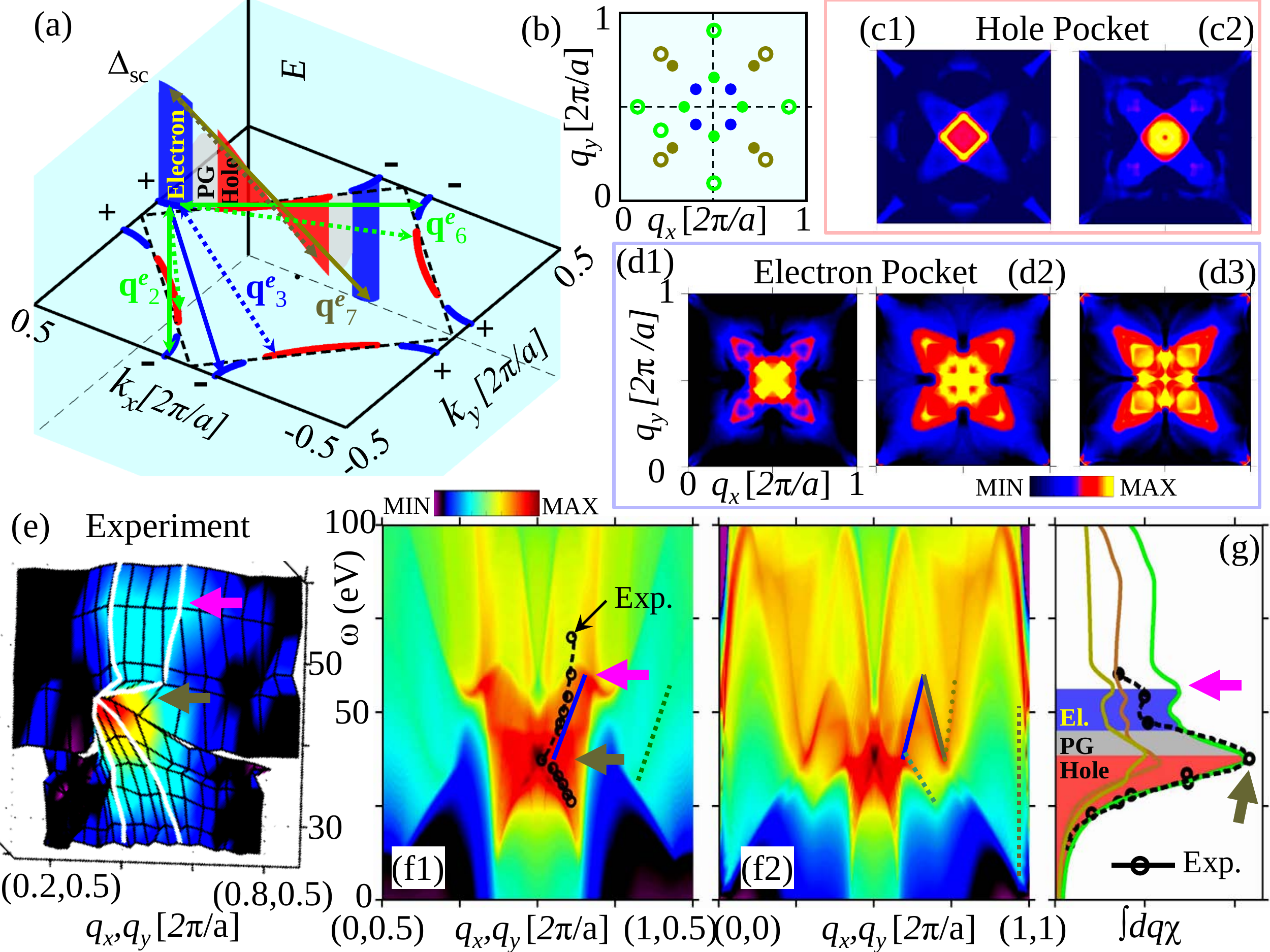}}}
\caption{{Magnetic resonance behavior in the electron-pocket.} (a) Schematic representation of the inelastic scattering process of Bogolyubov quasiparticles on the electron pockets. The out-of-plane red and blue shadings along the energy axis gives the superconducting gaps with $d-$wave symmetry. The solid arrows of same colors as in Figure~1 represent the same scattering vectors but here in the particle-hole channel. The dashed lines of same color are the same scattering channels but from the electron pocket to the hole-pocket and vice versa. (b) The scattering pattern expected at an energy corresponding to the electron-pocket. (c), The computed INS spectrum in the hole-pocket region is shown below the first resonance in (c1) and at the resonance in (c2).  (d) The spectra at three energy cuts above the first resonance in the electron-pocket region. (e) The INS data of YBCO$_y$ along (100)-direction at the same doping $y=6.6$ at which quantum oscillation measurements predict electron pockets. The magenta and gold arrows point to the two resonances coming from the hole-pocket and electron pocket, respectively. (f), Computed magnetic resonance spectra along (100)-direction and diagonal direction in momentum space, respectively. Solid and dashed lines of different colors are guides to the eye for different scattering branches, coming from scattering between electron-electron pocket and electron-hole pocket respectively. The dots are the experimental data, extracted by tracing the peak positions in the constant energy cuts of Neutron spectra shown in (f1). (g) The momentum integrated resonance intensities are shown for integration along (100)-direction (cyan), along the diagonal (gold), and the total (black). The computed results agree well with the experimental data for the same sample.} \label{fig3}
\end{figure*}

{\bf Inelastic Neutron Scattering Spectroscopy-}We turn next to the low-energy INS spectra in Fig.~\ref{fig3}, mainly in the region below $\omega \le2\Delta$ where Bogoliubov quasiparticle scattering dominates in the spin-excitation dispersion.\cite{chubukov,norman,eremin} INS measures the imaginary part of the susceptibility whose non-interacting part is $\chi_0^{\prime\prime}({\bm q},\omega_p)=\sum_{{\bm k},n}{\rm Im}\big[G({\bm k},i\omega_n)G({\bm k}+{\bm q},i\omega_n+\omega_p)\big]$, where $n$ is the Matsubara frequency index, see Ref.~\onlinecite{Dasresonance}. In the SC state, $\chi_0^{\prime\prime}$ arises from the inelastic scattering of the Cooper pairs (many body effects which are incorporated in the random-phase approximation shift the energy scale of the spectra to a slightly lower energy; nevertheless the overall shape of the spectrum is not greatly changed). Therefore, the spectrum is dominated by scattering by bright spots, similar to QPI but connecting features above and below the Fermi level. Among 7 ${\bm q}^{h,e}$ vectors in the QPI pattern discussed above only four vectors participate in the INS spectra, see Fig.~\ref{fig3}(a). Furthermore, owing to the selection rule associated with elastic scatterings in STM, ${\bm q}^h$ and ${\bm q}^e$ are always energy resolved. But in the INS spectra the separation between the two energy scales becomes obscured due to the turning on of inter pocket inelastic scattering. We denote the corresponding electron to hole pocket scattering channel by ${\bm q}^{eh}$ as shown by dashed lines of the same color in Fig.~\ref{fig3}(a). The resulting constant energy INS profile in the SC region is sketched in Fig.~\ref{fig3}(b).

In addition to the energy and momentum conservation principles associated with the inelastic Bogoliubov quasiparticle scattering, the coherence factors of both the superconducting state and SDW state play a major role here.\cite{chubukov,Dasresonance} The sign change of the superconducting order at the `hot-spot' ${\bm q}$, i.e. $\Delta_{k}=-\Delta_{k+q}$, is a crucial for finding non-vanishing contributions to the INS spectra. SDW order with a modulation vector $Q=(\pi,\pi)$ provides an additional coherence factor which leads to a gradual increase of intensity of the INS spectra as ${\bm q}$ approaches $Q$.

In the hole pocket region, our calculation correctly reproduces the magnetic resonance peak at $(\omega_{res}^h,Q)$ (magenta arrows in Figs.~\ref{fig3}(e),~\ref{fig3}(f) and \ref{fig3}(g) and both the downward and upward dispersions of the `hour-glass' pattern.\cite{chubukov,norman,eremin,Dasresonance} Below the resonance, the magnetic scattering of the Cooper pairs also yields the maximum intensity in the bond direction, see Fig.~\ref{fig3}(c1). In the absence of the electron-pocket, the INS intensity maxima rotate by 45$^o$ towards the diagonal direction above the resonance energy, again consistent with the hourglass phenomenology.

In the presence of the electron pocket, the INS pattern exhibits several distinguishing characteristics which can be separated from the usual hourglass pattern of the hole-pocket: (1) The intensity profile in the constant energy surface hosts peaks both along the bond direction as well as along the diagonal direction above the $(\pi,\pi)$-resonance, see Figs.~\ref{fig3}(d1), \ref{fig3}(d2) and \ref{fig3}(d3). (2) An additional resonance energy $\omega_{res}^e>\omega_{res}^h$ is observed along the bond direction in Fig.~3(f1), and in the integrated INS intensity in Fig.~\ref{fig3}(g). The presence of two resonances is also theoretically calculated for iron-pnictide superconductors, although the the differences in the FS topology and the pairing symmetry between these two classes of superconductors make the details of the resonance spectra look very different.\cite{Dastworesonance} (3) More resonance branches appear in the INS spectra although weak in intensity, in Figs.~\ref{fig3}(f1),~\ref{fig3}(f2).

The experimental results of YBCO$_{6.6}$ shows clear evidence for the second peak as shown in Figs.~\ref{fig3}(e) and \ref{fig3}(g). The energy scale of both the resonances are set by the SC gap amplitude as $\omega_{res}=2|\Delta_0g_{k}|$ where $g_k=[\cos{k_xa}-\cos{k_ya}]/2$ is the structure factor of the $d_{x^2-y^2}-$wave pairing. We have used the ARPES value of SC gap magnitude $\Delta_0=30$~meV from Refs.~\cite{JLee,Hueffner}. The two energy scales are determined by the position of the corresponding `hot-spot' momentum value on the FS. The first resonance occurs at $Q$ where the ${\bm q}_3$ vector connects the `bright-spots' at the tip of the hole pocket which gives $\omega_{h}=40$meV. The second resonance occurs when ${\bm q}_2$ touches the Brillouin zone boundary which yields $\omega_{res}^e=55$meV. Note that in our calculation, the two spin resonance energies have a direct relation to the peaks in the QPI, shown in Fig.~\ref{fig2}(c), where the spin resonance peak occurs at twice the energy of the peak in the QPI intensity.

\begin{figure}[top]
\hspace{-0cm}
\rotatebox{0}{\scalebox{.6}{\includegraphics{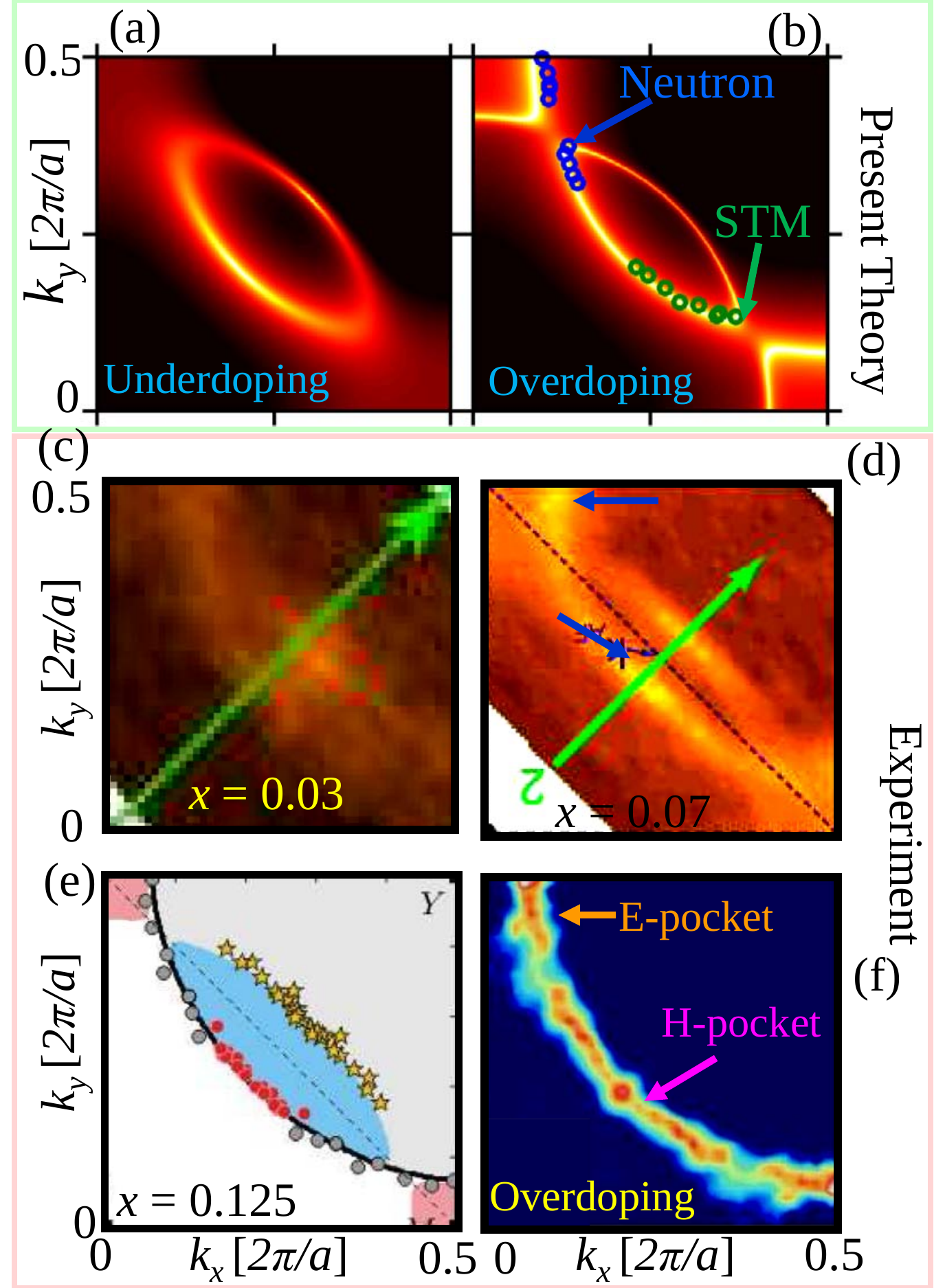}}}
\caption{{ARPES observation of electron pocket.} (a)-(b) Computed single particle spectral weight in the normal state at the Fermi level which gives the impression of a FS. In (a), only a hole-pocket is present at the nodal point while both the electron and hole-pocket are present in (b). All the calculations in the present manuscript are performed for the FS in (b). We extract the FS information from the experimental data of STM and INS presented in Figures~2 and 3, respectively which are plotted as open circles on top of the theory. (c)-(f) The experimental data of Fermi surface as a function of hole-dopings and material. The presence of the electron-pocket in (d)-(f), at the antinodal point can be identified by comparing the same with (c) which hosts only a hole-pocket. The data in (c)-(d) is obtained from LSCO,\cite{ARPESLSCO} while (e) is taken from Na-LSCO\cite{ARPESLBCO} and (f) is for an overdoped TBCO sample\cite{ARPESTBCO}. } \label{fig4}
\end{figure}

{\bf Angle-resolved Photoemission Spectroscopy-} The same information on the presence of the electron pocket can be directly obtained from ARPES. ARPES measures the single particle spectral weight $A({\bm k},\omega)=-{\rm Im}G({\bm k},\omega)/\pi$. In Fig.~\ref{fig4}, we provide some evidence for the presence of the electron pockets in the ARPES data.

In the strongly underdoped cuprates where the pseudogap is large, it gaps out the whole antinodal region above the magnetic zone boundary. Thus electron pockets disappear from the FS and only the hole-pocket is present, as shown in Fig.~\ref{fig4}(a). It is interesting to notice that even in the theoretical spectra, there is a finite incoherent spectral weight present away from the hole-pocket which traces the underlying ungapped FS. This is the effect of the imaginary part of the self-energy correction which is calculated to be quasi-linear in the low-energy region. As a result the residual spectral weight gradually decreases from the nodal to the antinodal regions as the pseudogap grows along the same direction but spreads over a larger energy and momentum region. This result is consistent with experiment in a deeply underdoped sample of LSCO as shown in Fig.~\ref{fig4}(c).

Therefore, in order to identify the electron pocket at the antinodal point, one needs to pay attention to the spectral weight. In the near-optimal region close to the quantum critical point, the electron pocket appears at the Fermi level, leading to coherent spectral weight at the antinodal point as shown in Fig.~\ref{fig4}(b). Looking at the experimental data for dopings $x=0.07-0.125$ of LSCO in Figs.~\ref{fig4}(d)-\ref{fig4}(e) and in an overdoped sample of TBCO in Fig.~\ref{fig4}(f), we see that both the hole and the electron pockets are present in this doping range. Especially, the spectral weight maps of the FS in Figs.~\ref{fig4}(d) and \ref{fig4}(f) have peaks of comparable magnitude at both nodal point and antinodal point while it is suppressed in between these two points. In the case of an ungapped full FS, the spectral weight is expected to be coherent and similar at each Fermi momentum, whereas as discussed above, when only a hole pocket is present the spectral weight gradually decreases from the nodal point to the antinodal point. Therefore, the experimental results in Figs.~\ref{fig4}(d)-\ref{fig4}(f) convincingly establish the presence of the electron pocket in the vicinity of  optimal doping for these two materials.

The procedure of inverting the QPI data to reconstruct the single-particle FS is well known\cite{Dasresonance} and following the conventional procedure, we find that the FS constructed from the experimental data of QPI maps used in Fig.~\ref{fig2} lies reasonably on top of the theoretical data. Note that the existing experimental data has not yet been analyzed with the notion to identify the electron pocket. Similarly, we extract the FS from the INS data of YBCO shown in Fig.~\ref{fig3} and the result agrees well with the picture of coexisting hole and electron pockets as shown in Fig.~\ref{fig4}(b).

Finally we comment on the difficulties of ARPES to observe the electron pocket. As mentioned earlier, electron-pockets are expected near the quantum critical point of the pseudogap at which the SC gap still survives. In this doping region, a typical phase diagram shows that the pseudogap transition temperature $T^*\lesssim T_c$ for most of the cuprates.\cite{Hueffner} Furthermore, in the SC state, the electron pockets, being positioned at the antinodal point are fully gapped by $d-$wave superconductivity and therefore, ARPES can not detect it. When temperature is increased above $T_c$ to close the SC gap, the small pseudogaps are also nearly closed, so the hole pocket and electron pocket disappears, and a full metallic FS forms. On the other hand, quantum oscillations are performed in high magnetic fields at which superconductivity is suppressed, and the electron pocket becomes exposed. Additional complications can arise since ARPES and STM are sensitive to the surface states as well as to so-called `matrix-element' effects, which could also explain failure to see certain portions of the FS at particular experimental conditions.

In conclusion, we present the detailed spectroscopic analysis of the electron pocket that will allow both single particle (ARPES) and two particle spectroscopies (STM and INS) to detect electron pockets that are posited to be present near optimal doping. These simple qualitative features provide a sharp contrast to a simple hole pocket models and hence offer a direct test of their presence. The simplest model that has electron pockets is the SDW state with or without coexisting  superconducting order. Even with this simplified model we find significant spectroscopic features that allow qualitative and quantitative determination of the electron pockets in cuprates.

\begin{appendix}

\section{Spin-density wave model for electron and hole pocket formation}\label{App1}

We use the tight-binding parametrization of our first-principles band structure of the antibonding band created by hybridization between Cu $d_{x^2-y^2}$ and O $p$ orbitals as our starting point. The FS reconstruction is modelled due to SDW which coexists with the $d_{x^2-y^2}-$wave superconductivity.\cite{Dastwogapcuprate} While we choose a $(\pi,\pi)-$modulation of the SDW, the results are general and are reproduced by charge density wave, $d$-density wave, or flux phase with similar modulation.\cite{Dastwogapcuprate} Furthermore, a two-dimensional stripe model with incommensurate modulation along the diagonal direction also predicts the coexistence of electron and hole-pockets in addition to other open FSs.\cite{Norman_stripe,Sachdev_Nernst,Harrison} Our obtained results of the QPI, Neutron and ARPES spectra are equivalent and reproducible as long as an electron pocket is present at $(\pi,0)/(0,\pi)-$points irrespective of its microscopic origin.

The Hubbard-BCS Hamiltonian in momentum space is\cite{Dastwogapcuprate}
\begin{eqnarray}
H &=& \sum_{{\bm k},\sigma}\xi_{\bm k}c^{\dag}_{{\bm k},\sigma}c_{{\bm k},\sigma}+U\sum_{{\bm k},{\bm k}^{\prime}} c^{\dag}_{{\bm k}+{\bm q},\uparrow}c_{{\bm k},\uparrow}c^{\dag}_{{\bm k}^{\prime}-{\bm q},\downarrow}c_{{\bm k}^{\prime},\downarrow}\nonumber\\
&&~~~~~~~~~~~+\sum_{{\bm k}}\Delta_{{\bm k}}c^{\dag}_{{\bm k},\uparrow}c^{\dag}_{-{\bm k},\downarrow}.
\label{Eq1}
\end{eqnarray}
where $c^{\dag}_{{\bm k},\sigma} (c_{{\bm k},\sigma}) $ is the electronic creation (destruction) operator with momentum ${\bm k}$ and spin $\sigma=\pm$.  $\xi_{\bm k}$ is the free particle dispersion. $U$ is the Hubbard onsite interaction term chosen to be 0.86eV. The SDW order parameter $S=\langle\sum_{{\bm k},\sigma}\sigma c^{\dag}_{{\bm k}+{\bm q},\sigma}c_{{\bm k},\sigma}\rangle=0.08$ is treated within self-consistent mean-field approximation. The BCS superconducting gap is $\Delta_{{\bm k}}=\Delta_0[\cos{(k_xa)}-\cos{(k_ya)}]/2.$, where $\Delta_0$=30meV is the experimental gap parameter for YBCO$_{6.6}$ taken from ARPES data.\cite{JLee,Hueffner}

The corresponding single particle $4\times 4$ Green's function is constructed from Eq.~\ref{Eq1} which includes Umklapp part from spin density wave and anomalous term coming from the SC gap.\cite{Dastwogapcuprate} The QPI maps and non-interacting susceptibility are calculated as convolutions of the Green's function $B({\bm q},\omega)=\sum_{\bm k}{\rm Im}\bigl[G({\bm k},\omega)G({\bm k}+{\bm q},\omega)\bigr]_{11}$ and $\chi({\bm q},\omega_p)=\sum_{\bm k}G({\bm k},i\omega_n)G({\bm k}+{\bm q},i\omega_n+\omega_p)$.

We calculate the self-energy due to all components of fluctuations along the spin and charge degrees of freedom within self-consistent GW-approach\cite{Dasop}
\begin{eqnarray}
\Sigma({\bm k},\sigma,i\omega_n)&=&\sum_{{\bm q},\sigma^{\prime}}\int_0^{\infty}\frac{d\omega_p}{2\pi}G({\bm k}+{\bm q},\sigma^{\prime},i\omega_n+\omega_p)\nonumber \\
&&\times\Gamma({\bm k},{\bm q},\omega,\omega_p)W({\bm q},\omega_p).
\end{eqnarray}
$W$ is the fluctuation potential obtained within random-phase approximation (RPA) as $1/2\eta U\chi_{RPA}^{\prime\prime}$, where $\chi_{RPA}^{\prime\prime}$ is the imaginary part of the RPA susceptibility of transverse spin ($\eta=2$), longitudinal spin ($\eta=1$) and charge ($\eta=1$) correlations functions. Finally, the vertex correction is approximated within Ward's identity as $\Gamma=(1-\partial\Sigma^{\prime}/\partial\omega)_0$. The calculation is performed in real frequency space using analytical continuation $i\omega_n\rightarrow\omega+i\delta$.

\begin{figure}[top]
\hspace{-0cm}
\rotatebox{0}{\scalebox{.6}{\includegraphics{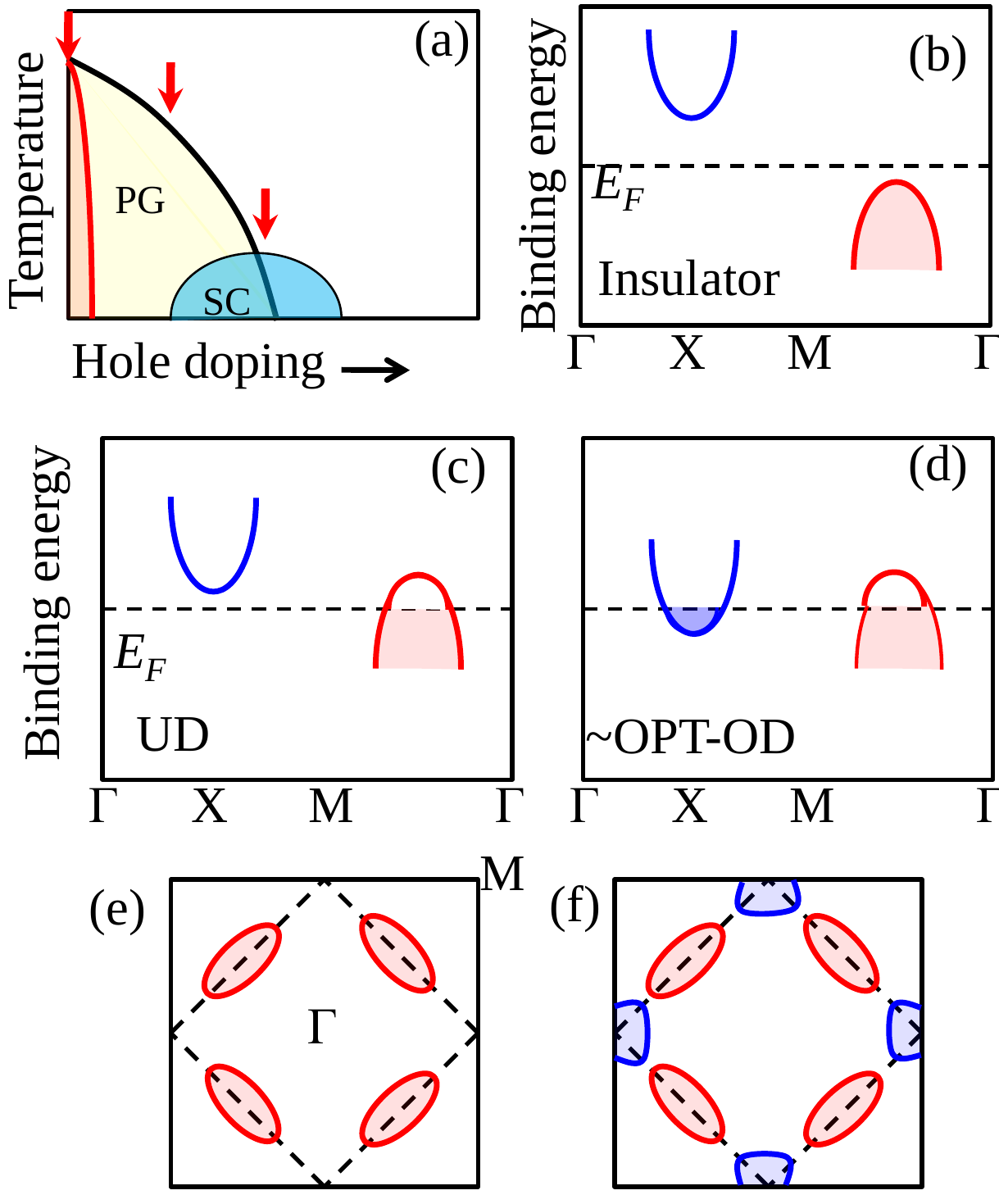}}}
\caption{{Schematic evolution of hole pocket to electron+hole pocket.} (a) Schematic phase diagram of hole doped cuprates. The detail of the relative doping dependence of the pseudogap and the superconducting gap is material specific. (b) SDW induced dispersion at half-filling. (b)-(c) The dispersions at finite dopings which lead to Fermi surfaces given in (e), (f) respectively.
(b) Hole pocket Fermi surface in underdoped cuprates. (c) Hole+electron pockets which constitute the Fermi surface near optimal doping.  } \label{fig5}
\end{figure}

Fig.~\ref{fig5} schematically demonstrates the evolution of the electron pocket as a function of doping in hole doped cuprates. At half-filling, strong SDW order opens up an insulating gap as shown in Fig.~\ref{fig5}(b). Doping reduces the strength of the pseudgap interaction, see Fig.~\ref{fig5}(a). With hole doping, the doped holes accumulate at the top of the lower SDW band [red line in Fig.~\ref{fig5}(c)], which give rise to a hole pocket at the nodel point as shown in Fig.~\ref{fig5}(e). Near optimal doping where the pseudogap is very small, the bottom of the upper SDW band drops below the Fermi level around ${\bf k}=(\pm\pi/a,0)/(0,\pm\pi/a)$ in Fig.~\ref{fig5}(d). Thus an electron pocket appears as shown in Fig.~\ref{fig5}(f). In this doping range the pseudogap opening shifts its location to the `hot-spot' region between the hole pocket and electron pocket. It should be noted that the magnitude of the pseudogap can be so small in this region that it may be overlooked due to the large superconducting gap for materials like Bi2212 or YBCO. With sufficiently large magnetic field when the superconducting gap is suppressed, the electron pocket becomes visible in quantum oscillation or Hall effect probes.

\subsection{QPI calculation}\label{AppQPI}
STM measures local density of states which is Fourier transformed into momentum space to obtain QPI maps. The local density of states in response to a local scalar scattering potential is defined as
\begin{eqnarray}
\rho({\bf r},{\bf r},\omega)&=&\sum_{{\bf r}_1}{\rm Im}\Bigl[G({\bf r},{\bf r}_1,\omega)V({\bf r}_1)G({\bf r}_1,{\bf r},\omega)\Bigr]\\
&=&\sum_{{\bf k},{\bf k}^{\prime}}{\rm Im}\Bigl[G({\bf k},\omega)G({\bf k}^{\prime},\omega)\nonumber\\
&&~~~~~~\times\sum_{{\bf r}_1}e^{i{\bf k}.({\bf r}-{\bf r}_1)}V({\bf r}_1)e^{i{\bf k}^{\prime}.({\bf r}_1-{\bf r})}\Bigr]\nonumber\\
%
%
&=&\sum_{{\bf k},{\bf q}}{\rm Im}\Bigl[G({\bf k},\omega)G({\bf k}+{\bf q},\omega)e^{i{\bf q}\cdot{\bf r}}V({\bf q })\Bigr].\nonumber\\
&=&\sum_{{\bf k},{\bf q}}V({\bf q })\Bigl[{\rm Im}\bigl[G({\bf k},\omega)G({\bf k}+{\bf q},\omega)\bigr]\cos{({\bf q}.{\bf r})} \nonumber\\
&&~~~+{\rm Re}\bigl[G({\bf k},\omega)G({\bf k}+{\bf q},\omega)\bigr]\sin{({\bf q}\cdot{\bf r})}\Bigr].\nonumber\\
\end{eqnarray}
Here ${\bf q}={\bf k}-{\bf k}^{\prime}$. In the case of an onsite potential, $V({\bf q })=V$. Finally, we take the Fourier transformation of the local density of states to obtain
\begin{eqnarray}
B({\bf q},\omega)&=&\sum_{\bf r}e^{i{\bf q}\cdot{\bf r}}\rho({\bf r},\omega)\\
%
&=&V\sum_{\bf r}(\cos{({\bf q}\cdot{\bf r})}+i\sin{({\bf q}\cdot{\bf r})})\nonumber\\
&&~~~\times\sum_{{\bf k},{\bf q}^{\prime}}\Bigl[{\rm Im}\bigl[G({\bf k},\omega)G({\bf k}+{\bf q}^{\prime},\omega)\bigr]\cos{({\bf q}^{\prime}\cdot{\bf r})} \nonumber\\
&&~~~+{\rm Re}\bigl[G({\bf k},\omega)G({\bf k}+{\bf q}^{\prime},\omega)\bigr]\sin{({\bf q}^{\prime}\cdot{\bf r})}\Bigr]\\
&\approx&V\sum_{{\bf k}}\Big[{\rm Im}\bigl[G({\bf k},\omega)G({\bf k}+{\bf q},\omega)\bigr]\nonumber\\
&&~~~~~~+i{\rm Re}\bigl[G({\bf k},\omega)G({\bf k}+{\bf q},\omega)\bigr]\Big].
\end{eqnarray}
In the above equation, we have incorporated the local field approximation which implies $\sum_{{\bf r},{\bf q}^{\prime}}\cos{({\bf q}\cdot{\bf r})}\cos{({\bf q}^{\prime}.{\bf r})}=\delta_{{\bf q},{\bf q}^{\prime}}$ and $\sum_{{\bf r},{\bf q}^{\prime}}\cos{({\bf q}.{\bf r})}\sin{({\bf q}^{\prime}.{\bf r})}=0$.
The summation is carried out over the entire reciprocal space but relaxing the Umklapp scattering condition to mimic the experimental procedure.\cite{Kohsaka}
\section{Comparing QPI maps for paramagnetic state with the only-hole pocket and hole+electron pocket states}\label{App3}
\begin{figure*}[top]
\hspace{-0cm}
\rotatebox{0}{\scalebox{.7}{\includegraphics{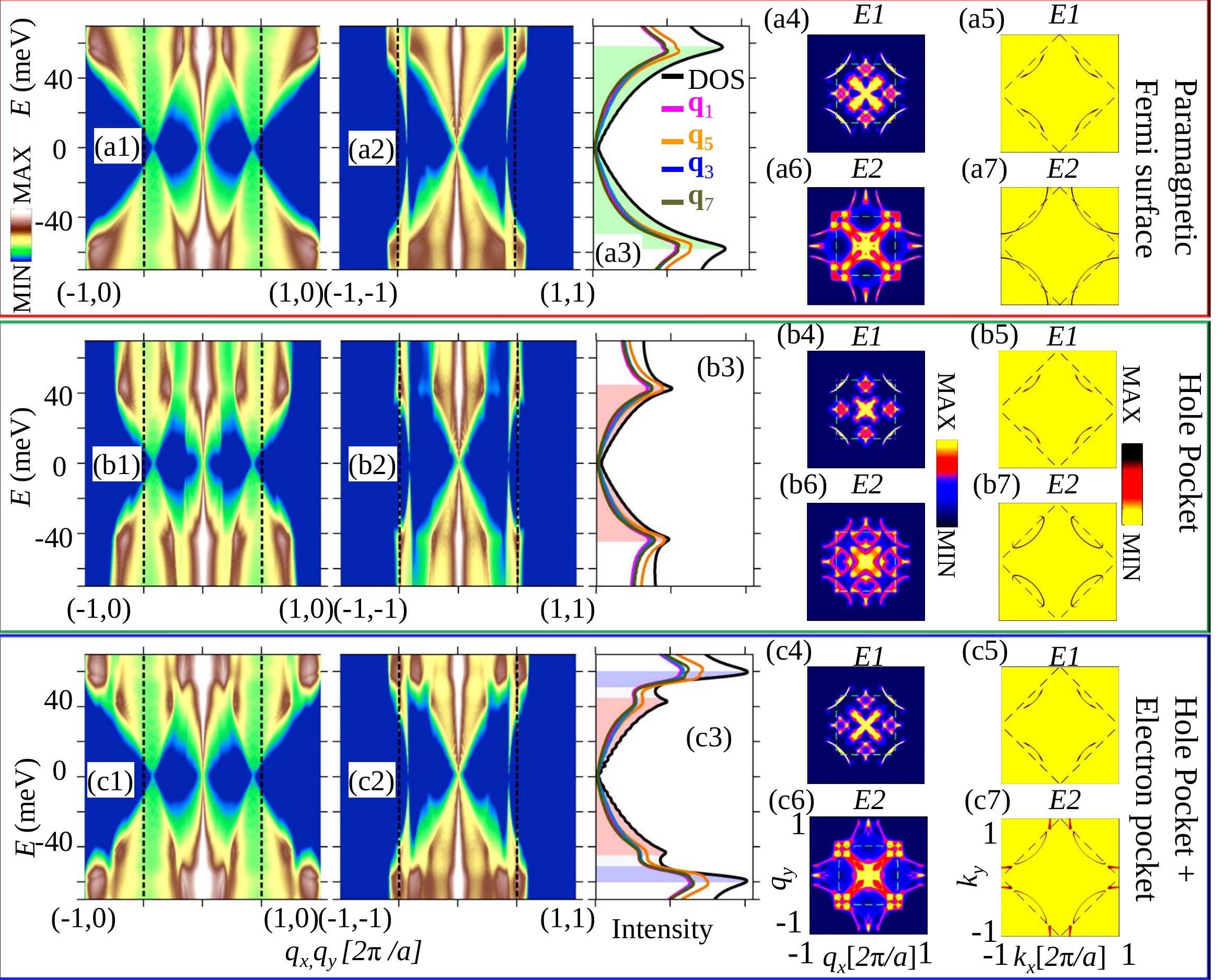}}}
\caption{{Comparison between the QPI patterns of paramagnetic full FS, hole-pocket FS, and coexisting electron and hole pocket FSs.} Three horizontal panels separated by boxes of different colors give the evolution of QPI patterns at the same energy and momentum cuts but for three different FS topologies. (a1), (b1), and (c1) are plotted along the (100) directions of the QPI profile whereas (a1), (b2) and (c2) are the same but along the diagonal directions. The QPI patterns and the corresponding constant energy `bright-spot' profiles are plotted at two different energy cuts in the third and fourth columns. The energy values $E1=25$meV and $E2=55$meV are kept same for all three cases for ease in comparison. $E1$ corresponds to the FS below the magnetic zone boundary (dashed green line in the last column) which is the hole pocket in the pseudogap state. $E2$ corresponds to an energy at which the `bright-spots' reside above the magnetic zone boundary for paramagnetic FS and electron pocket case and at the tip of the hole pocket for the middle panel. In all three cases, the superconducting parameters are kept constant while only the pseudogap strength is varied artificially to produce different FS topologies at the same doping.} \label{fig6}
\end{figure*}

In Fig.~\ref{fig6}, we compare the evolution of the QPI patterns in the case of a paramagnetic ground state, hole-pocket, and coexisting electron+hole pockets. As mentioned in the main text, there are several distinguishing features to unambiguously identify the electron pocket that will show up collectively in the dispersion, intensity, and constant energy profile of the QPI pattern. (1) In a paramagnetic state, all QPI vectors show continuous energy dependence with no 'kink' or non-dispersive pattern. For the case of a hole pocket without any electron pocket on the FS, all the dispersion features stop at the energy where the `bright-spots' reach to the top of the dispersion. No new QPI vector appears above this energy and along the $(100)$-direction, the QPI vectors do not extend to $(2\pi,0)$ while along the diagonal it does not cross the $(\pi,\pi)$-boundary.  On the other hand, in the case of coexisting electron and hole pockets both these features should be present. (2) The associated intensity of all QPI vectors also reflects the presence of an electron pocket. In a hole-pocket, a peak in the intensity occurs at the tip of the hole-pocket which is less than the SC gap amplitude. The peak extends to the SC gap amplitude in the case of a paramagnetic ground state. For the electron and hole pocket, both peaks will be present and can be used to identify the presence of an electron-pocket.  (3) Finally, the constant energy cuts of the QPI pattern can help distinguish the electron pocket from a hole-pocket, but the former can not be separated from a paramagnetic ground state. As discussed in point (1) above, if ${\bm q}_3>(\pi,\pi)$ as well as ${\bm q}_5=(2\pi,\pi)$ at some energy, that will be an unambiguous signature of the presence of an antinodal FS.

\end{appendix}

\begin{acknowledgments}
This work was funded by US DOE, BES and LDRD and benefited from the allocation of supercomputer time at NERSC.
\end{acknowledgments}

\end{document}